\documentclass[useAMS,usenatbib,usegraphicx,latexsym]{mn2e}
\usepackage{latexsym}

\usepackage{color}

\title[The 3.3$\,\umu$m PAH emission band of the Red Rectangle]{The 3.3$\,\umu$m PAH emission band of the Red Rectangle}

\author[I.-O. Song, J. McCombie, T.H. Kerr and P.J. Sarre]{In-Ok Song$^{1,2}$, J. McCombie$^{1}$, T. H. Kerr$^{3}$ and P. J. Sarre$^{1}$\\
$^{1}$School of Chemistry, The University of Nottingham, University Park, Nottingham NG7 2RD, U.K.\\
$^{2}$Department of Astronomy and Space Science, Kyung Hee University, Suwon, 446-701, Korea.\\
$^{3}$United Kingdom Infrared Telescope, Joint Astronomy Centre, 660
N. A'ohoku Place, University Park, Hilo, Hawaii 96720, U.S.A.}

\begin{document}

\date{Accepted. Received.}

\pagerange{\pageref{firstpage}--\pageref{lastpage}} \pubyear{2006}

\maketitle

\label{firstpage}

\begin{abstract}
A new analysis of long-slit CGS4 (UKIRT) spectra of the 3.3$\,\umu$m feature of the Red Rectangle and its evolution with offset along the NW
whisker of the nebula is presented. The results support a proposed two-component interpretation for the 3.3$\,\umu$m feature with peak wavelengths near 3.28$\,\umu$m and 3.30$\,\umu$m. Both components exhibit a small shift to shorter wavelength with increasing offset from the central star which, by comparison with laboratory
studies, is consistent with a decrease in temperature of the carriers with distance from HD~44179. The two-component approach is also applied to
3.3$\,\umu$m data for the Red Rectangle, Orion Bar D2 and Orion Bar H2S1 from ISO SWS studies.
\end{abstract}


\section{Introduction}

The unidentified infrared (UIR) bands are
a family of emission features at 3.3, 6.2, 7.7, 8.6, 11.2 and
12.7$\,\umu$m which are accompanied by several minor ones between 3 and
20$\,\umu$m. They are believed to be due to C-H and C-C vibrational transitions
of polycyclic aromatic hydrocarbon (PAH) molecules or small particles and have been observed in planetary and reflection nebulae, H\,\textsc{ii} regions, young stellar objects, the general
interstellar medium and galaxies such as ultraluminous infrared galaxies, star forming galaxies and active galactic nuclei.

Given the ubiquity
of the bands in both Galactic sources and external galaxies and their
widespread use as a probe of astronomical environments, improvement
in understanding their origin is a high priority. For recent discussion of the UIR bands see \citet{tok97,tie99,van00,sel01,hon01,pee02a,van04}.
Considerable observational, theoretical and
laboratory effort is being committed to PAH research and to the chemistry and physics of disks
and nebular material of objects such as the Red Rectangle, particularly as this holds the prospect of
providing much insight into the r\^{o}le of large molecules and dust
in astronomy in general.  It is notable that apart from
the related monocyclic benzene molecule to which a narrow absorption line at
14.84~$\umu$m in CRL 618 has been assigned \citep{cer01} and blue
fluorescence attributed to small PAHs in the Red Rectangle
\citep{vij04,vij05a}, there is no spectroscopic identification of an
individual PAH in any astronomical environment.

The 3.3$\,\umu$m feature is the shortest wavelength
member of the UIR band family and is normally attributed to the C-H
in-plane stretch vibration of PAHs.  Most researchers consider that it arises from a
superposition of infrared transitions of a number of gas-phase PAH
species following absorption of ultraviolet/visible photons, although
this is not the only proposed form of  PAH material or proposed
 excitation mechanism. Its profile has been explored in various objects
\citep{tok91,van04} and in some studies it is spatially resolved \citep{slo97,geb89,sel90, ker99, son03,gee05}.
These studies have suggested that there is no link between
the local radiation field and the profile of the 3.3$\,\umu$m
feature, but rather that the variation observed arises from compositional changes related to the age of
the emitting material \citep{sel90, van04}.
However, a clear picture for the material composition has not yet emerged and a coherent model remains to be established.

Of all currently studied Galactic objects the Red Rectangle is one of the most fascinating for UIR band studies.  It has a wealth of unidentified or partially identified emission features which fall in a wide spectral range from the near-IR to the far-UV \citep{coh75}. It is a particularly attractive target because of the strength of the UIR bands and its well-defined biconical geometry that extends at least 40$''$ from the centre of the nebula \citep{coh04}.
A new 3.28$\,\umu$m sub-feature was suggested
by \citet{son03} in order to interpret the 3.3$\,\umu$m band profile in the Red Rectangle.

In this paper we review briefly previous spectroscopic and imaging studies of the 3.3\,$\umu$m emission feature (section~2), summarize observational aspects (section~3), and present a two-component analysis of long-slit CGS4 3.3$\,\umu$m data for the Red Rectangle in section~4 with application also to ISO SWS 3.3\,$\umu$m profiles of the Orion Bar (section~5). Discussion and summary/conclusions are given in sections~6 and 7, respectively.

\section{The C-H emission feature: a brief review of observations}

\subsection{Spectroscopy and profile of the 3.3$\,\umu$m band}

The origin and nature of the 3.3$\,\umu$m band has been discussed by
a number of authors \citep{sel90,tok91,van04}. The profile is known to vary
between and within objects and an (\textbf{A, B}) classification has
been put forward by van Diedenhoven et al. (2004) with
particular reference to ISO SWS data. The \textbf{A$_{3.3}$} group
is common and has a profile described as `symmetric' with peak position at $\sim$3.290$\,\umu$m and a FWHM of 0.040$\,\umu$m.  This profile is found in widely differing objects ranging from Orion Bar H2S1 to NGC~7027 and
 is similar to Type 1 ($\lambda_{max}$$\sim$3.289$\,\umu$m; FWHM~$\le$~0.042$\,\umu$m) in the earlier classification of
\citet{tok91}. Group \textbf{B1$_{3.3}$} with maximum at
$\sim$3.293$\,\umu$m (FWHM $\sim$0.037$\,\umu$m) and group
\textbf{B2$_{3.3}$} with a peak wavelength at $\sim$3.297$\,\umu$m
(FWHM $\sim$0.037$\,\umu$m) are found in relatively few objects, an example of \textbf{B1$_{3.3}$} being Orion Bar~D2.

The overall Red Rectangle spectrum is classified as
\textbf{B2$_{3.3}$} where this comprises a 14$''$$\times$20$''$
exposure of both star and nebula.  The 3.3$\,\umu$m
profile directly on the star HD~44179 is
fitted very well by a single Lorentzian function (Song et al. 2003); it is of the rarer Type 2 in the \citet{tok91} description.  Hence the ISO SWS (\textbf{B2$_{3.3}$}) Red Rectangle spectrum is a superposition of Type~2 and 1 where the latter persists in the nebula.   Song et al. (2003) showed that the development of a short-wavelength
shoulder on the 3.3$\,\umu$m feature with increasing offset from the
central star could be interpreted in terms of the growth of a new
band with a peak wavelength of 3.28$\,\umu$m. This represents the evolution from a Type~2 to Type~1 (\textbf{A$_{3.3}$}) profile.

We describe in this paper a refined
analysis of the spatially resolved Red Rectangle data of \citet{son03} which supports the existence
of the proposed 3.28$\,\umu$m band and establishes small
blueward shifts in peak wavelength with offset for both components
of the overall 3.3$\,\umu$m feature.  The same approach is applied to some ISO SWS targets in section 5. We
use the nominal description `3.3'$\,\umu$m to refer to the whole
3.3$\,\umu$m profile irrespective of shape, with the
wavelengths of the components written with a second decimal place
\emph{viz} 3.28$\,\umu$m and 3.30$\,\umu$m.

\subsection{Imaging and spatially-resolved studies}
The binary star \citep{wae96} of the Red Rectangle is obscured by a
disk that lies close to the W-E axis \citep{rod95,ost97, tut02,wat98}.
The \emph{inner} region of $\le$\,1$''$ has been the
subject of high-spatial-resolution imaging infrared studies
\citep{rod95,ost97,men98,men02,mek98,tut02,mur03,miy04} and radio
mapping in CO \citep{buj05}. In the study of \citet{son03} the slit
was positioned (a) directly on HD~44179 and (b) aligned
along the NW interface but offset by 2$''$ from the star.

The spatial distribution of IR emission in both the 3.3$\,\umu$m
emission feature and the continuum for a \emph{wider} region of
the nebula has been investigated using both imaging and
long-slit spectroscopic techniques. \citet{bre93a} found the
continuum-subtracted 3.3$\,\umu$m distribution to be centrally
peaked on the star and slightly extended N-S over a region up to
\emph{c.} 4$''$ from the source at a spatial resolution of
$\sim$0.5-1$''$. At a higher spatial resolution of $\sim$0.2$''$,
\citet{mek98} reported more detail in the 3.3$\,\umu$m distribution
and found elongation along the cone walls within 1-1.5$''$ of the
star about the N-S axis. Spectroscopic
measurements by \citet{ker99} with a slit in four positions placed
5$''$ N, S, E and W of the star revealed 3.3$\,\umu$m PAH emission
nearly symmetrically distributed around the star at these higher
offsets. Significantly there was no enhancement of 3.3$\,\umu$m emission at the
intersections of the slit with the bicone interfaces where the
unidentified optical emission features peak in intensity.

The distribution of the 3.3$\,\umu$m feature contrasts sharply with images taken in other C-H mode UIR bands. \citet{bre93a} found that the
11.3$\,\umu$m image shows a N-S bipolar shape with no central peak. Similar images have been reported by \citet{hor96} for a
5$''$~$\times$~5$''$ region at the C-H-related UIR wavelengths of 8.6, 11.2 and 12.7$\,\umu$m, but not for the continuum wavelengths of 10.0 and
20.2$\,\umu$m which are centrally peaked. From these  results it would appear that although the 3.3$\,\umu$m feature is a PAH transition,
it does not originate in the same class of PAHs that give rise to the C-H UIR bands at longer wavelengths. A spatial distinction
between 3.3$\,\umu$m and other PAH modes has also been found in NGC~1333 and the Orion bar and attributed to differences in the size of the
carriers, with the 3.3$\,\umu$m band arising in smaller PAHs \citep{bre93b, bre94}.

\section[]{Observational details}

\subsection{UKIRT CGS4 observations of the Red Rectangle}
The technical details and a log of the near-IR long-slit observations have
been presented by \citet{son03}. As this paper describes a new
analysis of these data, we summarize here that they were recorded
using the 1-5$\,\umu$m CGS4 spectrometer with a 1-pixel wide slit
(0.6$''$)and a 256~$\times$~256 InSb array giving a resolving power of 1000 at 3.3$\,\umu$m. Spectra were taken with the 90$''$ long
slit aligned along the NW `whisker'. Except for recording an on-star
spectrum, the slit was offset from the central star by 2$''$ in
order to avoid saturation of the CCD by the flux of the star.

\subsection{ISO SWS observations}
ISO data for these objects were taken from \citet{van04} and used in their published form.  The spectra were obtained with the SWS
(de Graauw et al. 1996) with a resolving power of 500-1500, processed using the package IA3, and rebinned with a constant resolution as described in \citet{pee02b}.

\begin{table}
\begin{flushleft}
\caption{\label{table:BandQuantity} Intensity, peak wavelength and
FWHM of the overall 3.3$\,\umu$m band as a function of offset from
HD~44179 along the NW interface of the nebula. Columns 3 and 4 are
the values obtained from the best (single) Lorentzian fit to the
asymmetric profile (see text).} \vspace{0.2cm}
  {
  \renewcommand{\arraystretch}{.9}
  \footnotesize
\begin{center}
\begin{tabular}{cccc}
\hline\hline
Offset/    & Intensity/ & Peak Wavelength/ & FWHM/ \\
arcsec & 10$^{-25}$Wcm$^{-2}$ & $\umu$m (${\rm cm^{-1}}$) & $\umu$m (${\rm cm^{-1}}$) \\
\hline
2.00 & 379.58 &3.297  (3033) & 0.036  (33) \\
2.61 & 266.89 &3.297  (3033) & 0.035  (32) \\
3.22 & 60.54  &3.297  (3033) & 0.034  (31) \\
3.83 & 11.01  &3.295  (3035) & 0.034  (32) \\
4.44 & 4.01   &3.293  (3036) & 0.036  (33) \\
5.05 & 2.02   &3.293  (3037) & 0.037  (34) \\
5.66 & 1.17   &3.293  (3037) & 0.038  (35) \\
6.27 & 0.72 &3.292  (3037) & 0.039  (36) \\
6.88 & 0.51 &3.291  (3038) & 0.040  (37) \\
7.49 & 0.38 &3.291  (3038) & 0.041  (38) \\
8.10 & 0.27 &3.292  (3038) & 0.041  (37) \\
8.71 & 0.19 &3.291  (3039) & 0.042  (39) \\
9.32 & 0.14 &3.290  (3040) & 0.043  (40) \\
9.93 & 0.13  &3.290  (3040) & 0.050  (46) \\
10.54 & 0.10 &3.290 (3039) & 0.045  (42) \\
\hline\hline
\end{tabular}
\end{center}
  }
\end{flushleft}
\end{table}

\begin{figure}
\includegraphics[width=84mm]{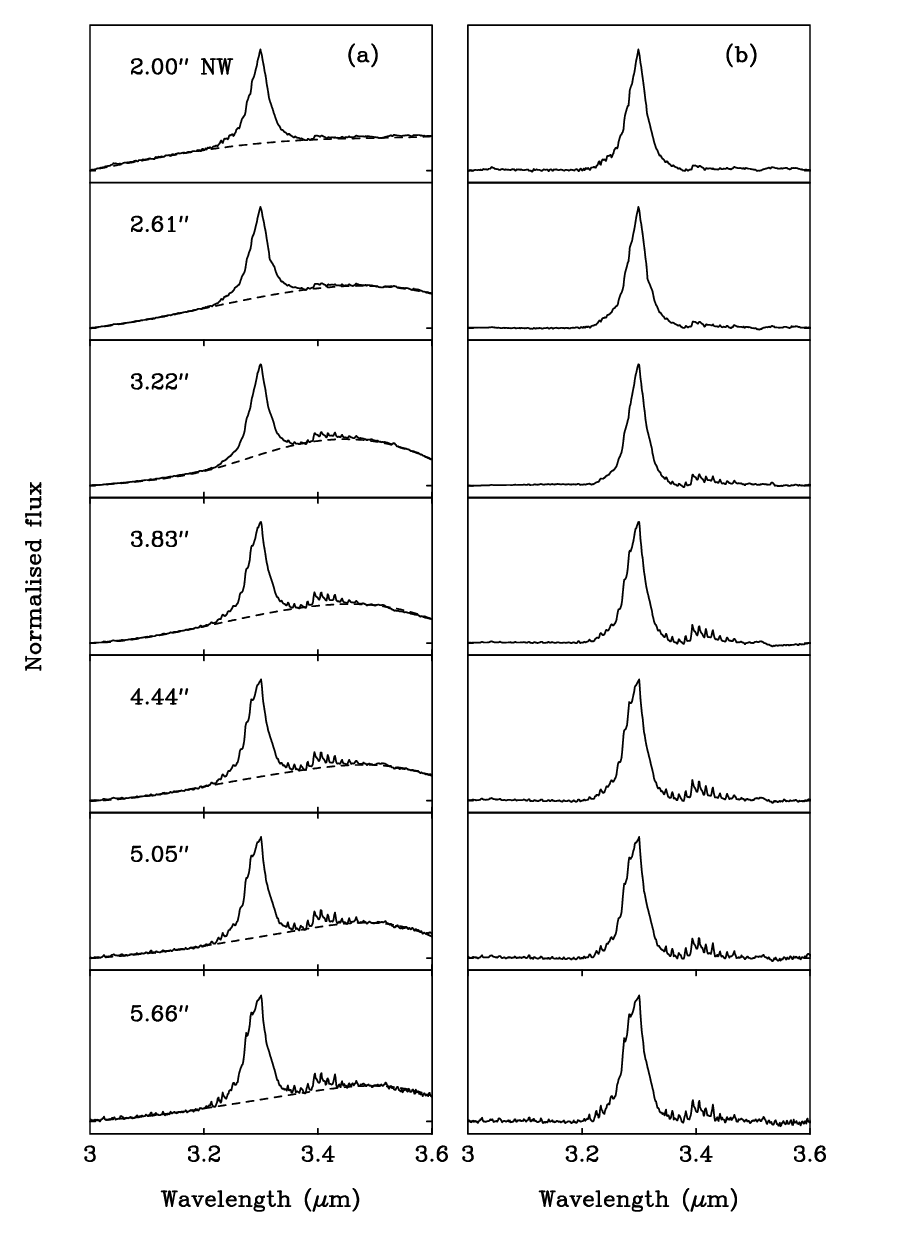}
\caption{Spatially resolved spectra (solid line) of the 3.3$\,\umu$m feature 2-6$''$ along the NW interface of the Red Rectangle with (a) a polynomial fit to the continuum (dashed line), and (b) continuum-subtracted spectra.}
\end{figure}

\begin{figure}
\includegraphics[width=84mm]{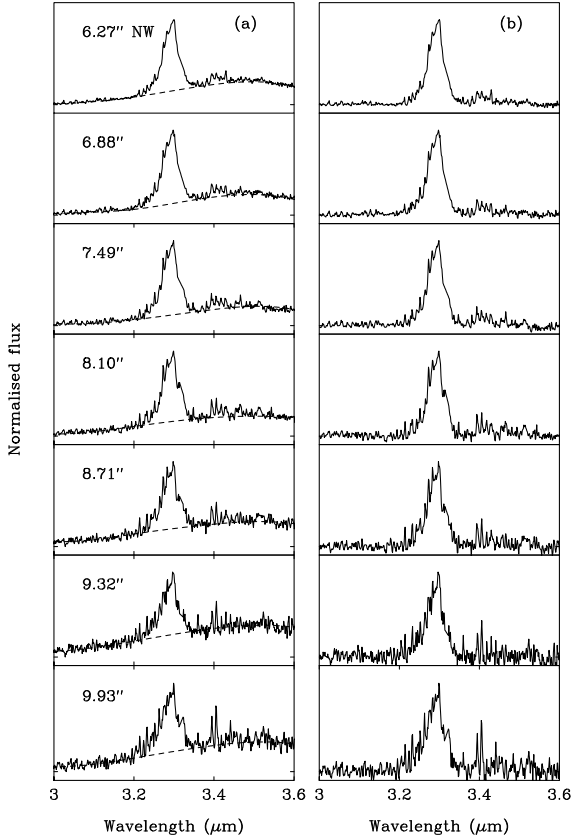}
\caption{Spatially resolved spectra (solid line) of the 3.3$\,\umu$m feature 6-10$''$ along the NW interface of the Red Rectangle with (a) a polynomial fit to the continuum (dashed line), and (b) continuum-subtracted spectra.} \label{fig2}
\end{figure}


\begin{figure}
\includegraphics[width=82mm]{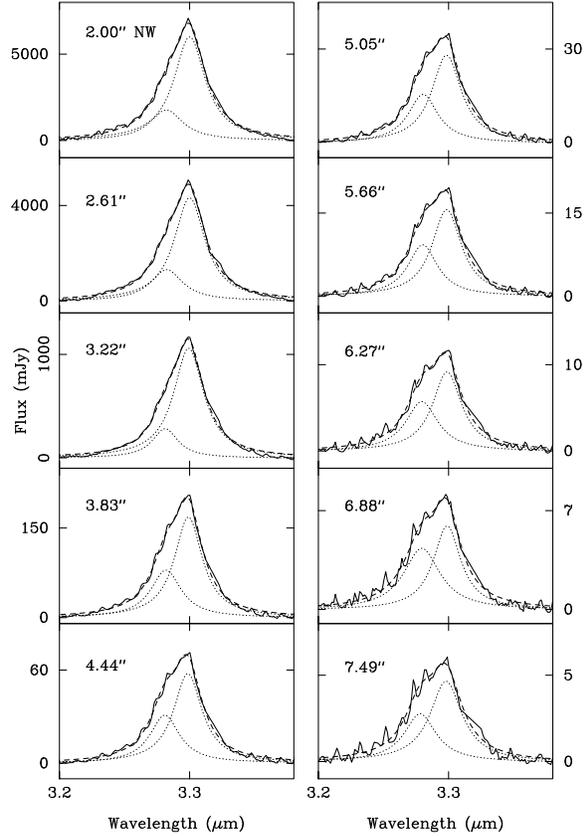}
\caption{Fitted Lorentzian profiles as a function of offset from the central star. The solid lines are the observed
  spectra, the dotted lines are individual Lorentzian profiles
  and the dashed lines are the sum of the two profiles. The
  3.28$\,\umu$m/3.30$\,\umu$m intensity ratio increases with offset.} \label{fig3}
\end{figure}

\begin{figure}
\includegraphics[width=84mm]{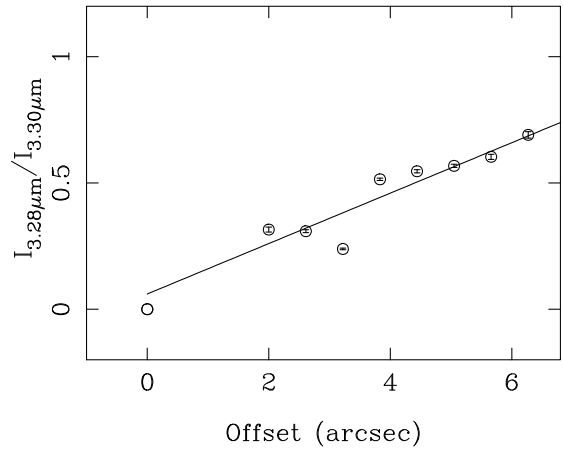}
\caption{Plot of the 3.28$\,\umu$m/3.30$\,\umu$m intensity ratio as a function of offset from HD~44179 along the NW whisker of the Red Rectangle.  The line is the result of an unweighted linear least-squares fit to the data above 2$''$.  There are no data points between 0 and 2$''$ because the slit was offset from the star.} \label{fig4}
\end{figure}

\begin{figure}
\includegraphics[width=84mm]{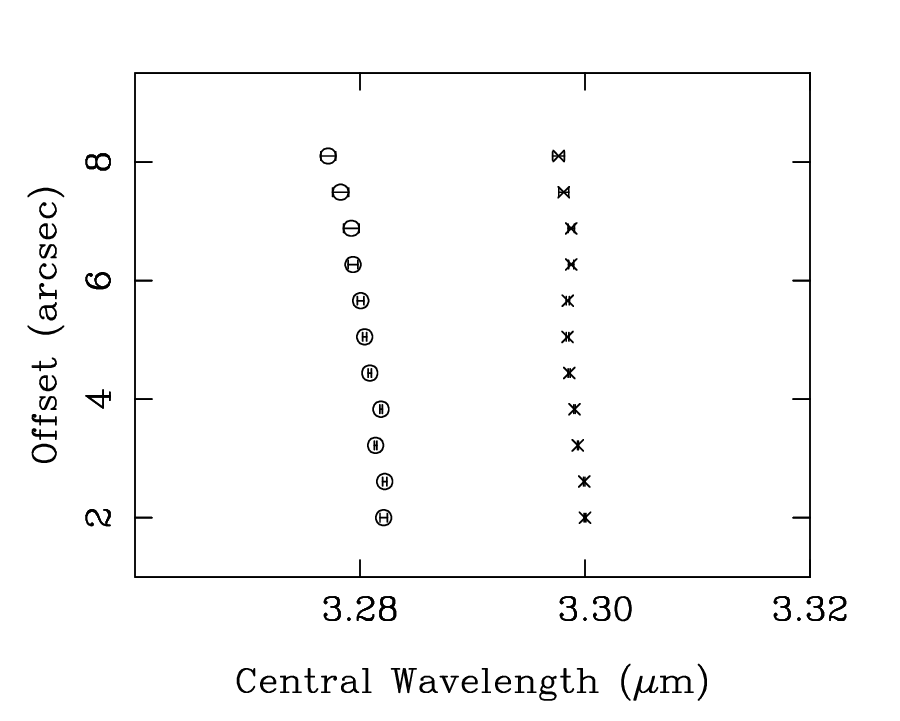}
\caption{Plot of the central wavelength of the 3.28 (o) and 3.30$\,\umu$m (x) components as a function of offset in the Red Rectangle.  The values were obtained by
Lorentzian fitting of the overall 3.3$\,\umu$m profile (see text).} \label{fig10}
\end{figure}

\begin{figure}
\includegraphics[width=84mm]{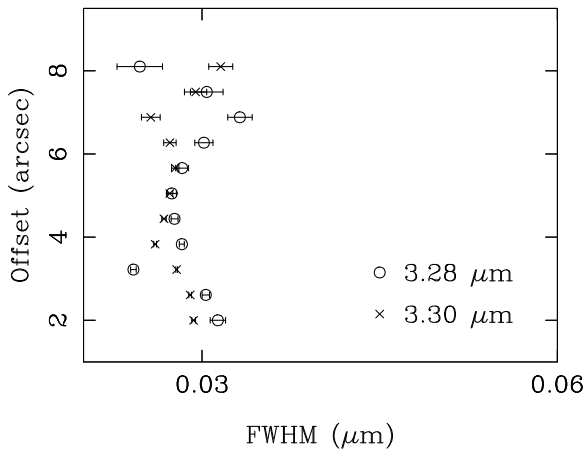}
\caption{Plot of the width of the 3.28 (o) and 3.30$\,\umu$m (x)
components as a function of offset in the Red Rectangle. The values were obtained by
Lorentzian fitting of the overall 3.3$\,\umu$m profile (see text).} \label{fig6}
\end{figure}

\begin{figure}
\includegraphics[width=84mm]{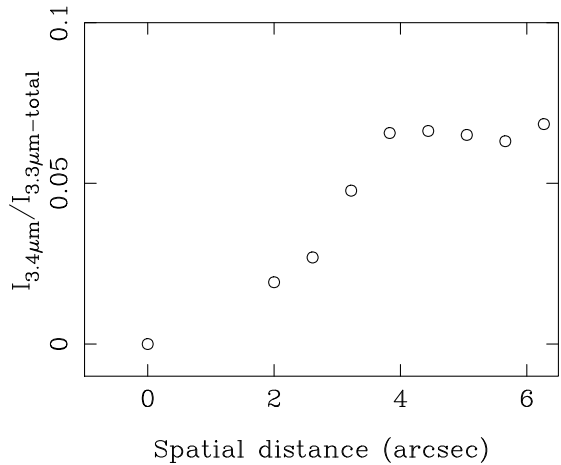}
\caption{3.4$\,\umu$m/3.3$\,\umu$m (integrated) intensity ratio along the
NW whisker of the Red Rectangle. Residual telluric contamination in the 
3.4$\,\umu$m region makes it difficult to estimate the errors in this 
case.}\label{fig7}
\end{figure}

\section{Red Rectangle and ISO SWS spectra}

\subsection{Two-component analysis of the 3.3$\,\umu$m profile of the Red Rectangle}

The overall 3.3$\,\umu$m band maximum shifts to shorter wavelength,
increases in width and declines rapidly in intensity with offset
from HD~44179 along the NW interface. This is summarized in table 1 where a single
Lorentzian fit to the 3.3$\,\umu$m band was employed to illustrate these characteristics.
However, the band shape in the nebula is
\emph{asymmetric} and varies as shown for offsets between 2$''$ and
10$''$ in figures 1 and 2. \citet{son03} suggested that the evolution of
the profile could be interpreted in terms of the emergence of a new
3.28$\,\umu$m band which gave the overall appearance of a shoulder
on the short-wavelength side. There is no evidence for a 3.28$\,\umu$m component in the on-star spectrum which is
symmetrical and very well fitted by a single Lorentzian with peak
maximum at 3.298$\,\umu$m and FWHM~=~32 cm$^{-1}$ ($\sim$0.035$\,\umu$m) as shown in fig 1.
of \citet{son03}. The 3.3$\,\umu$m band in the nebula was previously analysed
simply by taking the fit to the on-star spectrum and subtracting
this from the recorded profiles at the various distances into the
nebula. Here we employ an improved approach in which the overall
3.3$\,\umu$m band is fitted with two Lorentzians, allowing the
intensities, peak wavelengths and widths to float in the fitting
procedure (\textbf{splot} in \textsc{iraf} -- which uses the Levenberg-Marquardt
optimisation).  The results of the new fitting for offsets between 2$''$
and 8$''$ along the NW axis are shown in fig. 3 where the
importance of the 3.28$\,\umu$m contribution is seen to grow with
offset.  The 3.28$\,\umu$m/3.30$\,\umu$m ratio grows approximately linearly with
distance as shown in fig. 4 and suggests that at very high offset
 (towards the ISM) the 3.28$\,\umu$m feature would dominate.  This is consistent with
 the observation of a 3.28$\,\umu$m absorption feature towards the
 Galactic centre \citep{chi00,son03}.  We consider possible origins of the 3.30 and 3.28$\,\umu$m components in the next section.

The peak maxima of the two components from the fit shift to slightly 
shorter wavelength with distance from HD~44179 (see fig. 5).  
The blueward
shifts of the peak wavelength between 2$''$ and 8$''$ are $\sim$2
cm$^{-1}$ and $\sim$4 cm$^{-1}$ for the 3.30 and 3.28$\,\umu$m
components, respectively (fig. 5). 
The widths 
of the components appear to show some variation with offset (see fig. 6) 
and, if confirmed, may be due to changes in the PAH size distribution with 
offset.
If the shift is caused by a reduction in
internal temperature of the carriers and if the carriers are indeed
gas-phase molecules, the direction of the shift is in the sense
expected in comparison with experimental and theoretical data for
gas-phase PAHs \citep{job95}. Laboratory studies of the 3.3$\,\umu$m
\emph{absorption} bands of naphthalene (C$_{10}$H$_8$), pyrene
(C$_{16}$H$_{10}$), coronene (C$_{24}$H$_{12}$) and ovalene
(C$_{32}$H$_{14}$) as a function of temperature show that the
temperature-dependent frequency shift for these molecules increases
with molecular size with values of \emph{c.} $-0.0139$, $-0.0284$, $-0.0328$ and
$-0.042$~cm$^{-1}$K$^{-1}$, respectively \citep{job95}. The largest of these laboratory
coefficients suggests a carrier temperature reduction of only
$\sim$100~K over 2-8$''$.  However, given that the infrared emission
 is thought to arise from rapid heating due to photon absorption rather than through thermal emission, this small temperature reduction is
probably acceptable. The lower frequency shift with offset for the
3.30$\,\umu$m component (the one most prominent on-star) suggests
that this feature arises in smaller PAHs than those that give rise to the 3.28$\,\umu$m band
which is (relatively) stronger in the nebula. The temperature
dependence of the width of the laboratory 3.3$\,\umu$m absorption feature on molecular size is
less pronounced ranging from 0.0353~cm$^{-1}$K$^{-1}$ for naphthalene
to 0.056~cm$^{-1}$K$^{-1}$ for ovalene \citep{job95}. This is
qualitatively consistent with the astrophysical observations.

\subsection{Origin of two 3.3$\,\umu$m components}

Possible interpretations for the existence of two components are that these arise from
different structural forms, sizes or hydrogenation/ionisation states
of the ensemble of UIR band emitters.  Laboratory gas-phase \emph{absorption}
spectra taken at elevated temperatures indicate that a blueward shift of the 3.3$\,\umu$m
peak maximum occurs with increasing PAH size of \emph{c. }
15 cm$^{-1}$ (0.015$\,\umu$m) from pyrene (C$_{16}$H$_{10}$) to ovalene
(C$_{32}$H$_{14}$) \citep{job95}.  This is of the same magnitude as the \emph{c.} 0.02$\,\umu$m difference between
the 3.30$\,\umu$m and 3.28$\,\umu$m features discussed here.  A wider range of gas-phase data
particularly for larger PAHs would be of much interest.  \citet{van04} comment that
the 3.3$\,\umu$m feature is attributable to the smallest emitting
PAHs \citep{ala89,sch93}, and report that within the NASA Ames
laboratory sample of neutral PAH spectra taken in an inert gas matrix the C-H stretch frequency
distribution is \emph{bimodal} with the bands of the smallest PAHs
centred near 3060 cm$^{-1}$ (3.27$\,\umu$m) and transitions of
larger PAHs occurring near $\sim$3090 cm$^{-1}$ (3.24$\,\umu$m). The
origin of this bimodal distribution is not clear. Applying a redward
shift of $\sim$0.03$\,\umu$m to these laboratory data, as is commonly invoked on
account of the higher carrier temperature in the nebula \citep{van04}, yields wavelengths of 3.30$\,\umu$m
($N_C \leq 40$) and 3.27$\,\umu$m ($N_C \geq 40$) where $N_C$ is the number of carbon atoms. These values are
in good agreement with the wavelengths of the two components
deduced in the analysis of \citet{son03} and considered further here.  Finally we note that although laboratory data show that the
3.3$\,\umu$m stretch frequency varies with PAH size and temperature, reported DFT
B3LYP/4-31G C-H stretching frequencies are virtually independent of
size being 3064, 3064 and 3062~cm$^{-1}$ for C$_{24}$H$_{12}$
(coronene), C$_{54}$H$_{18}$ (circumcoronene) and C$_{96}$H$_{24}$
\citep{bau02}.\\

\subsection{Comparison of 3.4$\,\umu$m and 3.28$\,\umu$m emission of the Red Rectangle}

As for the 3.28$\,\umu$m band, the 3.4$\,\umu$m emission is not present
on-star but grows in strength relative to the 3.3$\,\umu$m band with
offset. This emission almost certainly
arises from C-H motion of side groups such as -CH$_2$ and -CH$_3$ or from
doubly hydrogenated sites \citep{geb89,wag00,pau01}. Figure 7. shows a plot of the 3.4$\,\umu$m/3.3$\,\umu$m band ratio
as a function of offset where the values run from zero on-star
to $\sim$0.06 at 4-6$''$; the trend may be compared with that for the 3.28$\,\umu$m feature (figure 4). This (relative) growth in strength of the 3.4~$\,\umu$m band with offset is in the same sense as that found by \citet{geb89} who,
using a 5$''$ aperture, determined a 3.4$\,\umu$m/3.3$\,\umu$m ratio of 0.06 on-star
(which includes part of the nebula) and 0.13 at a position 5$''$~N. It is striking that the 3.28$\,\umu$m and 3.4$\,\umu$m bands exhibit a very similar behaviour with increasing offset.

\section{Application of two-component analysis to ISO SWS 3.3$\,\umu$m spectra}

We have also applied the two-component fitting approach to 3.3$\,\umu$m
ISO data from \citet{van04} for Orion Bar H2S1 and D2 (see figure 8),
and for the Red Rectangle (14$''$$\times$ 20$''$ \emph{i.e.} central
star plus nebula). The ISO 3.3$\,\umu$m spectra for these objects together with the long-slit data for the Red Rectangle (on-star and 3.8$''$ offset) are shown in figure 9. Not only does the extent of the blue shoulder vary between objects, but the long-wavelength side of the feature also shifts slightly.

The results of two-component fitting of the Orion Bar D2 and Orion Bar
H2S1 data are shown in figures 10 and 11 and may be compared with the
spatially resolved fits for the Red Rectangle in figure 3. The
spectral shape for Orion Bar D2 approximates to the higher offset
region of the Red Rectangle, whereas the Orion Bar H2S1 profile is broader and symmetric (Class \textbf{A$_{3.3}$}).

Figure 12. presents the 3.28$\,\umu$m/3.30$\,\umu$m integrated intensity ratios obtained
from Lorentzian fitting.  It includes the long-slit CGS4 3.3$\,\umu$m
`on-star' data ($\bullet$) and shows the behaviour of the 3.28$\,\umu$m/3.30$\,\umu$m ratio with
increasing offset into the Red Rectangle nebula. The on-star
profile is unique within this data set with no 3.28$\,\umu$m
component as noted previously \citep{son03}, so the ratio (y axis
value in figure 12) is zero. From this analysis Orion Bar H2S1 has
the highest 3.28$\,\umu$m/3.30$\,\umu$m ratio and exceeds unity, with Orion Bar D2 and RR ISO data falling close to that for the $\sim$8$''$ offset
position of the Red Rectangle from the spatially-resolved CGS4 long-slit observations.\\
\\
\begin{figure}
\includegraphics[width=84mm]{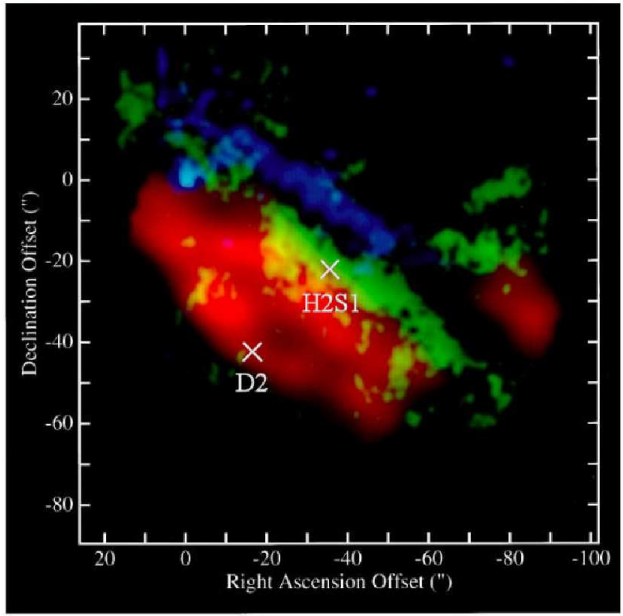}
\caption{Orion Bar PDR showing the distribution of 3.3$\,\umu$m PAH emission (blue), 2.12$\,\umu$m  H$_2$ v=1-0 S(1) emission (green) and $^{12}$CO J = 1-0 emission (red).  The star $\theta$$^2A$Ori is at the (0,0) position at R.A. = 05 35 22.5, Dec. = -05 24 57.8 (2000). Reproduced (and adapted) with permission from Tielens et al. (1993)} \label{fig8}
\end{figure}

\begin{figure}
\includegraphics[width=82mm]{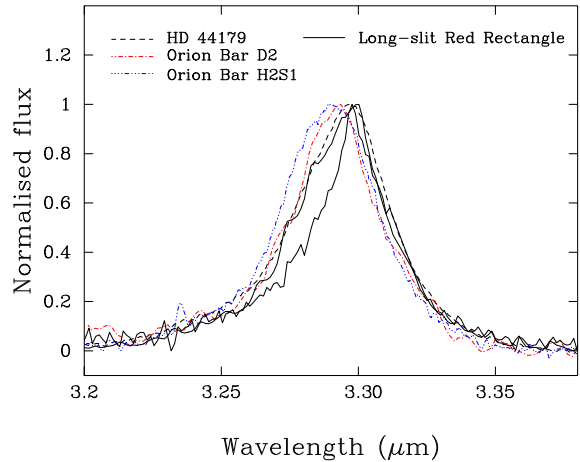}
\caption{Normalised 3.3$\,\umu$m profiles for three ISO SWS targets
and long-slit Red Rectangle data (on-star and 3.8$''$ offset).
The continuous lines represent the long-slit CGS4 Red Rectangle data, the inner
trace being the on-star spectrum.} \label{fig7}
\end{figure}

\begin{figure}
\includegraphics[width=84mm]{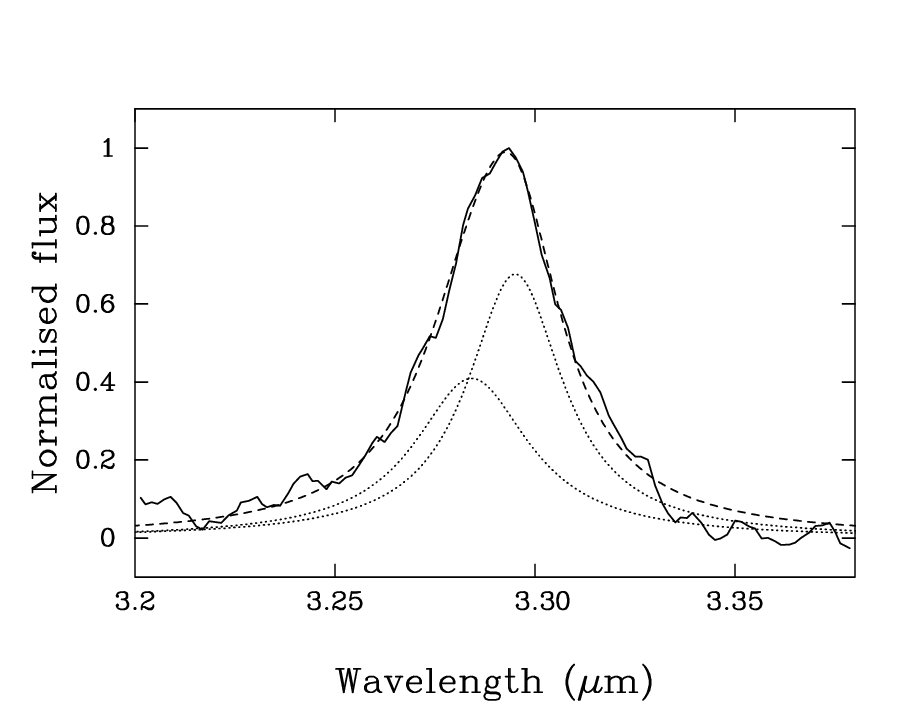}
\caption{Two-component Lorentzian fit to the 3.3$\,\umu$m profile
towards Orion Bar D2 as recorded with ISO SWS.} \label{fig8}
\end{figure}

\begin{figure}
\includegraphics[width=84mm]{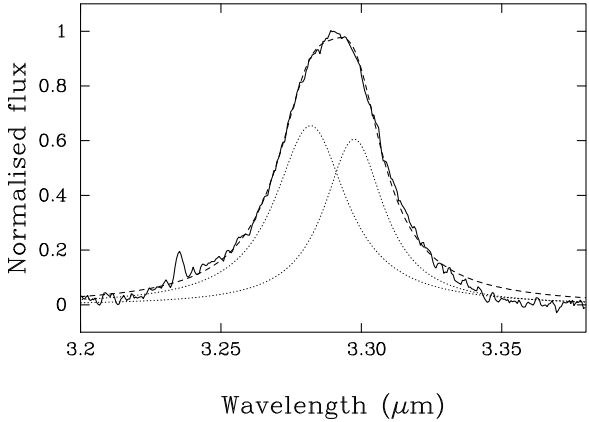}
\caption{Two-component Lorentzian fit to the 3.3$\,\umu$m profile
towards Orion Bar H2S1 as recorded by ISO SWS.} \label{fig9}
\end{figure}

\begin{figure}
\includegraphics[width=84mm]{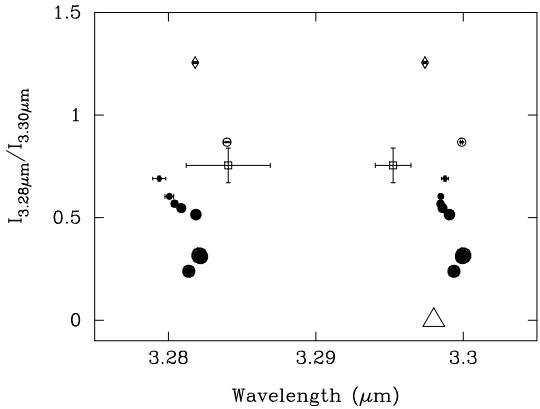}
\caption{Plot of 3.28$\,\umu$m/3.30$\,\umu$m intensity ratio \emph{vs.}
peak wavelength of the fitted two-component Lorentzian profiles. The symbols are: 
$\diamondsuit$ --- Orion Bar HS21, $\Box$ --- Orion Bar D2, $\circ$ --- Red Rectangle (ISO data), and $\bullet$ --- Red Rectangle (this work) where the diameter of the circle ($\bullet$) is inversely proportional to the offset from the star and the data extend in this figure to 6$''$.3. The open $\triangle$ indicates the on-star datum for which the 3.28$\,\umu$m feature is absent.} \label{fig10}
\end{figure}

\section{Discussion}
We suggest that the variation in profile of the 3.3$\,\umu$m PAH band is most readily interpreted in terms
 of the size distribution of neutral PAHs.  The HD~44179 (on-star) 3.30$\,\umu$m band is then characteristic
 of the smallest PAHs, with a larger PAH (3.28$\,\umu$m) population growing
 with offset in the Red Rectangle in parallel with growth of the 3.4$\,\umu$m complex due to addition of chemical
 groups. In comparison with ISO data for Orion Bar D2 and H2S1, the outer
 part of the Red Rectangle nebula has a 3.28$\,\umu$m/3.30$\,\umu$m ratio comparable to that of Orion Bar D2, but H2S1, which is
 nearer the ionisation front, has 3.28$\,\umu$m and 3.30$\,\umu$m components that
 are approximately equal in strength.

  This pattern appears to follow the level of extinction, $A_V$, having
 values of about 4.2 (Red Rectangle on-star though strongly dependent on geometry due to the disk \citep{men02,vij05a}), $A_V$~$>$~4 for
 Orion D2 \citep{tie93} and $A_V$~$<$~4 for Orion H2S1 \citep{tie93}.  This implies that only in regions of relatively high extinction can smaller PAHs survive as discussed by \citet{vij05b}.  This does not exclude the possibility of some larger PAHs also near HD~44179 because the observability of IR emission depends also on the intrinsic efficiency of conversion of UV to IR radiation which may well be greater for small PAHs. We also remark that the evolution of material from `fresh' near to HD~44179 to `processed' seems to follow the development of the 3.28$\,\umu$m feature.

\section{Summary and Conclusions}

A re-analysis of long-slit spectra using a
two-component fitting supports the earlier suggestion of two components of the 3.3$\,\umu$m band in the Red Rectangle nebula, centred near 3.30$\,\umu$m and 3.28$\,\umu$m.  Both components exhibit small shifts with offset attributable to a temperature decrease. On the basis of band shifts,
band ratios and comparison with laboratory data, the observations
are consistent with the smallest neutral PAHs in the Red Rectangle
being present `on'-star (with no 3.28$\,\umu$m component), with
the size of PAHs growing with offset and giving rise the 3.28$\,\umu$m band.

\section*{Acknowledgments}

We thank the UK Panel for the Allocation of Telescope Time for the
award of observing time on UKIRT, the National Institute for
International Education Department (NIIED) of the Korean Government
and The University of Nottingham for a studentship to In-Ok Song.  We are grateful to Els Peeters for making available to us the ISO SWS 3.3$\,\umu$m data in electronic form.

\label{lastpage}


\begin{thebibliography}{99}

\bibitem[\protect\citeauthoryear{Allamandola, Tielens \& Barker}{1989}]{ala89} Allamandola, L.J., Tielens, A.G.G.M., Barker, J.
R., 1989, ApJS, 71, 733

\bibitem[\protect\citeauthoryear{Bauschlicher}{2002}]{bau02}Bauschlicher, C.W., 2002, ApJ, 564, 782

\bibitem[\protect\citeauthoryear{Bregman et
al.}{1993a}]{bre93a} Bregman, J.D., Rank, D., Temi, P., Hudgins, D.,
Kay, L., 1993a, ApJ, 411, 794

\bibitem[\protect\citeauthoryear{Bregman et
al.}{1993b}]{bre93b} Bregman, J., Rank, D., Sandford, S.A., Temi,
P., 1993b, ApJ, 410, 668

\bibitem[\protect\citeauthoryear{Bregman et al.}{1994}]{bre94} Bregman, J., Larson, K., Rank, D.,
Temi, P., 1994, ApJ, 423, 326.

\bibitem[\protect\citeauthoryear{Bujarrabal et al.}{2005}]{buj05} Bujarrabal, V., Castro-Carrizo, A., Alcolea, J., Neri, R., 2005,
A\&A, 441, 1031 and references therein.


\bibitem[\protect\citeauthoryear{Cernicharo et al.}{2001}]{cer01} Cernicharo, J., Heras, A.M., Tielens, A.G.G.M., Pardo, J.R., Herpin, F., Gu\'{e}lin, M., Waters, L.B.F.M., 2001, ApJ, 546, L123


\bibitem[\protect\citeauthoryear{Chiar et al.}{2000}]{chi00} Chiar J.E., Tielens, A.G.G.M., Whittet, D.C.B., Schutte, W.A., Boogert, A.C.A., Lutz, D., van Dishoeck, E.F., Bernstein, M.P., 2000, ApJ, 537, 749



\bibitem[\protect\citeauthoryear{Cohen et al.}{1975}]{coh75} Cohen M. et al., 1975, ApJ, 196, 179

\bibitem[\protect\citeauthoryear{Cohen et al.}{2004}]{coh04} Cohen M., Van Winckel, H., Bond, H.E., Gull, T.R., 2004, AJ, 127, 2362

\bibitem[\protect\citeauthoryear{de Graauw et al.}{1996}]{deg96} de Graauw, T., et al., 1996, A\&A, 315, L49



\bibitem[\protect\citeauthoryear{Geers et
al.}{2005}]{gee05} Geers, V.C., Augereau, J., Pontoppidan, K.M.,
K\"{a}ufl, H., Lagrange, A., Chauvin, G., van Dishoeck, E.F., 2005,
High resolution infrared spectroscopy in astronomy, ed. H. U.
K\"{a}ufl, R. Siebenmorgen, and A.F.M. Moorwood. ESO astrophysics
symposia (Berlin: Springer), 239


\bibitem[\protect\citeauthoryear{Geballe et al.}{1989}]{geb89} Geballe, T.R., Tielens, A.G.G.M., Allamandola, L.J., Moorhouse, A., Brand, P.W.J.L., 1989, ApJ, 341, 278

\bibitem[\protect\citeauthoryear{Hony et al.}{2001}]{hon01} Hony, S., Van Kerckhoven, C., Peeters, E., Tielens, A.G.G.M., Hudgins, D.M., Allamandola, L.J., 2001, A\&A, 370, 1030

\bibitem[\protect\citeauthoryear{Hora et al.}{1996}]{hor96} Hora, J.L., Deutsch, L.K., Hoffmann, W.F., Fazio,
G.G., 1996, AJ, 112, 2064


\bibitem[\protect\citeauthoryear{Joblin et al.}{1995}]{job95} Joblin, C., Boissel, P., L\'{e}ger, A., D'Hendecourt, L., D\'{e}fourneau,
D., 1995, A\&A, 299, 835


\bibitem[\protect\citeauthoryear{Kerr et al.}{1999}]{ker99} Kerr T.H., Hurst M.E., Miles J.R., Sarre P.J., 1999, MNRAS, 303, 446



\bibitem[\protect\citeauthoryear{M\'{e}karnia et al.}{1998}]{mek98} M\'{e}karnia, D., Rouan, D., Tessier, E., Dougados, C., Lef\`{e}vre, J.,
1998, A\&A, 336, 648

\bibitem[\protect\citeauthoryear{Men'shchikov et al.}{1998}]{men98} Men'shchikov, A.B., Balega, Y.Y., Osterbart, G., Weigelt, G., 1998,
New Astron., 3, 601

\bibitem[\protect\citeauthoryear{Men'shchikov et al.}{2002}]{men02} Men'shchikov, A.B., Schertl, D., Tuthill, P.G., Weigelt, G., Yungelson, L.R., 2002, A\&A, 393, 867

\bibitem[\protect\citeauthoryear{Miyata et al.}{2004}]{miy04} Miyata, T., Kataza, H., Okamoto, Y. K., Onaka, T., Sako, S., Honda, M., Yamashita, T., Murakawa, K., 2004, A\&A, 415, 179

\bibitem[\protect\citeauthoryear{Murakawa et al.}{2003}]{mur03} Murakawa, K., Tamura, M., Suto, H., Miyata,
T., Gledhill, T.M., Hough, J.H.  2003, Proc. SPIE, 4843, 196

\bibitem[\protect\citeauthoryear{Osterbart, Langer \& Weigelt}{1997}]{ost97} Osterbart, R., Langer, N., Weigelt,
G., 1997, A\&A, 325, 609



\bibitem[\protect\citeauthoryear{Pauzat \& Ellinger}{2001}]{pau01} Pauzat, F., Ellinger, Y., 2001, MNRAS, 324,
355 and references therein.


\bibitem[\protect\citeauthoryear{Peeters et
al.}{2002a}]{pee02a} Peeters, E., Hony, S., Van Kerckhoven, C.,
Tielens, A.G.G.M., Allamandola, L.J., Hudgins, D.M., Bauschlicher,
C.W., 2002a, A\&A, 390, 1089

\bibitem[\protect\citeauthoryear{Peeters et al.}{2002b}]{pee02b} Peeters, E. et al., 2002b, A\&A, 381, 571


\bibitem[\protect\citeauthoryear{Roddier et al.}{1995}]{rod95} Roddier, F., Roddier, C., Graves, J. E., Northcott, M.
J., 1995, ApJ, 443, 249.


\bibitem[\protect\citeauthoryear{Schutte, Tielens \& Allamandola}{1993}]{sch93} Schutte, W.A., Tielens, A.G.G.M., Allamandola, L.J., 1993, ApJ, 415, 397

\bibitem[\protect\citeauthoryear{Sellgren, Tokunaga \& Nakada}{1990}]{sel90} Sellgren, K., Tokunaga, A.T., Nakada, Y., 1990, ApJ, 349, 120

\bibitem[\protect\citeauthoryear{Sellgren}{2001}]{sel01} Sellgren, K., 2001, Spectrochim. Acta A, 57, 627

\bibitem[\protect\citeauthoryear{Sloan et al.}{1997}]{slo97} Sloan, G.C., Bregman, J.D., Geballe, T.R., Allamandola, L.J., Woodward, C.E., 1997, ApJ, 474, 735

\bibitem[\protect\citeauthoryear{Song et al.}{2003}]{son03} Song, I.-O., Kerr, T.H., McCombie, J., Sarre, P.J., 2003, MNRAS, 346, L1


\bibitem[\protect\citeauthoryear{Tielens et al.}{1993}]{tie93} Tielens, A.G.G.M., Meixner, M.M., van der Werf, P.P., Bregman, J., Tauber, J.A., Stutzki, J., \& Rank, D., 1993, Science, 262, 86

\bibitem[\protect\citeauthoryear{Tielens et al.}{1999}]{tie99} Tielens, A.G.G.M., Hony, S., Van Kerckhoven, C., Peeters, E., 1999, in Proc. ESA Symp., The Universe as seen by \emph{ISO}, ed. P. Cox \& M.F. Kessler, (ESA SP-427; Noordwijk: ESA), 579

\bibitem[\protect\citeauthoryear{Tokunaga}{1997}]{tok97}
Tokunaga, A.T., 1997, in ASP Conf. Ser. 124, Diffuse Infrared
Radiation and the IRTS, ed. H. Okuda, T. Matsumoto, T. Roellig (San
Francisco: ASP), 149

\bibitem[\protect\citeauthoryear{Tokunaga et
al.}{1991}]{tok91} Tokunaga, A.T., Sellgren, K., Smith, R.G.,
Nagata, T., Sakata, A., Nakada, Y., 1991, ApJ, 380, 452


\bibitem[\protect\citeauthoryear{Tuthill et al.}{2002}]{tut02} Tuthill, P.G., Men'shchikov, A.B., Schertl, D., Monnier, J.D., Danchi, W.C.,
Weigelt, G., 2002, A\&A, 389, 889

\bibitem[\protect\citeauthoryear{van Diedenhoven et al.}{2004}]{van04} van Diedenhoven, B., Peeters, E., Van Kerckhoven, C., Hony, S., Hudgins, D.M., Allamandola, L.J., Tielens,
A.G.G.M., 2004, ApJ, 611, 928 and references therein.

\bibitem[\protect\citeauthoryear{Van Kerckhoven et al.}{2000}]{van00} Van Kerckhoven, C. et al., 2000, A\&A, 357, 1013


\bibitem[\protect\citeauthoryear{Vijh, Witt \& Gordon}{2004}]{vij04} Vijh, U.P., Witt, A.N., Gordon, K.D., 2004, ApJ, 606, L65

\bibitem[\protect\citeauthoryear{Vijh, Witt \&
 Gordon}{2005a}]{vij05a} Vijh, U.P., Witt, A.N., Gordon, K.D., 2005a,
ApJ, 619, 368
\bibitem[\protect\citeauthoryear{Vijh, Witt \&
Gordon}{2005b}]{vij05b} Vijh, U.P., Witt, A.N., Gordon, K.D., 2005b,
ApJ, 633, 262



\bibitem[\protect\citeauthoryear{Waelkens et al.}{1996}]{wae96} Waelkens C., Van Winckel H., Waters L.B.F.M., Bakker E.J., 1996, A\&A, 314, L17

\bibitem[\protect\citeauthoryear{Wagner, Kim \&
Saykally}{2000}]{wag00} Wagner, D.R., Kim, H.-S., Saykally, R.J.,
2000, ApJ, 545, 854

\bibitem[\protect\citeauthoryear{Waters et
al.}{1998}]{wat98} Waters L.B.F.M. et al., 1998, Nature, 391, 868



\end{thebibliography}
\end{document}